\documentclass[twocolumn]{aastex62}
\usepackage{amsmath}
\usepackage{booktabs}
\usepackage{fontawesome}
\usepackage{bm}
\usepackage{amsbsy}
\usepackage{fontawesome}

\usepackage{soul, CJK}
\usepackage[utf8]{inputenc}

\usepackage{xspace}
\DeclareFixedFont{\ttb}{T1}{txtt}{bx}{n}{9} 
\DeclareFixedFont{\ttm}{T1}{txtt}{m}{n}{9}  
\usepackage{color}
\definecolor{deepblue}{rgb}{0,0,0.5}
\definecolor{deepred}{rgb}{0.6,0,0}
\definecolor{deepgreen}{rgb}{0,0.5,0}
\usepackage{listings}
\usepackage{gensymb}

\newcommand\pythonstyle{\lstset{
language=Python,
basicstyle=\ttm,
otherkeywords={self},             
keywordstyle=\ttb\color{deepblue},
emph={MyClass,__init__},          
emphstyle=\ttb\color{deepred},    
stringstyle=\color{deepgreen},
frame=tb,                         
showstringspaces=false            %
}}
\lstnewenvironment{pyt}[1][]
{
\pythonstyle
\lstset{#1}
}
{}

\newcommand{\br}[1]{\mathopen{}\left(#1\right)\mathclose{}}

\newcommand{\bigo}[1]{\mathcal{O}\br{#1}}

\newcommand{\nul}{\xi_{\rm null}}
\newcommand{\nult}{\xi_{\rm null,0}}

\usepackage{listings}

\graphicspath{{./}{figures/}}

\shorttitle{A Mathematical Treatment of the Offset Degeneracy}
\shortauthors{Zhang \& Gaudi}

\begin{document}

\title{A Mathematical Treatment of the Offset Microlensing Degeneracy}

\correspondingauthor{Keming Zhang}
\email{kemingz@berkeley.edu}

\author[0000-0002-9870-5695]{Keming Zhang \begin{CJK*}{UTF8}{gkai}(张可名)\end{CJK*}}
\affil{Department of Astronomy, University of California, Berkeley, CA 94720-3411, USA}
\author[0000-0003-0395-9869]{B. Scott Gaudi}
\affil{Department of Astronomy, The Ohio State University, Columbus, OH 43210, USA}

\begin{abstract}

The offset microlensing degeneracy, recently proposed by \cite{zhang_ubiquitous_2022}, has been shown to generalize the close-wide and inner-outer \textit{caustic} degeneracies into a unified regime of \textit{magnification} degeneracy in the interpretation of 2-body planetary microlensing observations.
While the inner-outer degeneracy expects the source trajectory to pass equidistant to the planetary caustics of the degenerate lens configurations, the offset degeneracy states that the same mathematical expression applies to any combination of the close, wide, and resonant caustic topologies, where the projected star-planet separations differ by an offset ($s_{\rm A}\neq s_{\rm B}$) that depends on where the source trajectory crosses the star-planet axis.
An important implication is that the $s_{\rm A}=1/s_{\rm B}$ solution of the close-wide degeneracy never strictly manifests in observations except when the source crosses a singular point near the primary.
Nevertheless, the offset degeneracy was proposed upon numerical calculations, and no theoretical justification was given.
Here, we provide a theoretical treatment of the offset degeneracy, which demonstrates its nature as a mathematical degeneracy. From first principles, we show that the offset degeneracy formalism is exact to zeroth-order in the mass ratio ($q$) for two cases: when the source crosses the lens-axis inside of caustics, and for $(s_{\rm A}-s_{\rm B})^6\ll1$ when crossing outside of caustics. 
The extent to which the offset degeneracy persists in oblique source trajectories is explored numerically.
Lastly, it is shown that the superposition principle allows for a straightforward generalization to $N$-body microlenses with $N-1$ planetary lens components ($q\ll1$), which results in a $2^{N-1}$-fold degeneracy.

\end{abstract}

\keywords{Binary lens microlensing (2136), Gravitational microlensing exoplanet detection (2147)}

\section{Introduction}
\label{sec:intro}
Photometric observations of planetary microlensing events are commonly subject to a 2-fold-degenerate interpretation where the projected planet location differs ($s_A\neq s_B$) but the planet-to-star mass ratio remains the same ($q_A=q_B$). The close-wide degeneracy (e.g.\ \citealt{griest_use_1998,dominik_binary_1999,an_gravitational_2005}) is commonly invoked for such events with source stars passing close to the central caustic, while the inner-outer degeneracy \citep{gaudi_planet_1997,han_moa-2016-blg-319lb_2018} is cited for events which have source stars passing close to the planetary caustic. The close-wide degeneracy arises from the invariance of the shape and size of the central caustic under the $s\leftrightarrow 1/s$ transformation for $|1-s|\gg q^{1/3}$, a condition which is equivalent to the lens system being far from the resonant regime \citep{an_condition_2021}. The inner-outer degeneracy arises from the Chang-Refsdal \citep{chang_star_1984} approximation to the planetary caustics (\citealt{gaudi_planet_1997,dominik_binary_1999}), which describes a point-mass lens with uniform shear. Chang-Refsdal caustics are symmetric both along the star-planet axis (referred to as the lens axis hereafter), and along the line perpendicular to the star-planet axis that runs through the center of the caustic.

Recently, \cite{yee_ogle-2019-blg-0960_2021} and \cite{zhang_ubiquitous_2022} noted various inconsistencies of the two aforementioned degeneracies with those seen in real and simulated events. \cite{yee_ogle-2019-blg-0960_2021} noted the large number of semi-resonant topology events that cite the close-wide degeneracy, for which the degenerate solutions do not exactly follow $s\leftrightarrow 1/s$ nor satisfy $|1-s|\gg q^{1/3}$. They went on to suggest that there may be a continuum between the close-wide and inner-outer degeneracies in the resonant regime.
Subsequently, \cite{zhang_ubiquitous_2022} pointed out that the $s\leftrightarrow 1/s$ relationship is also not exactly followed even within the $|1-s|\gg q^{1/3}$ regime in which the close-wide degeneracy is expected to hold. They pointed out that the close-wide and inner-outer degeneracies are fundamentally {\it caustic} degeneracies which do not necessarily translate to {\it magnification} degeneracies that manifest in light-curves.

The offset degeneracy \citep{zhang_ubiquitous_2022} is then proposed independently of caustics as a magnification degeneracy, which both relaxes the non-resonant condition ($|1-s|\gg q^{1/3}$) and resolves the aforementioned inconsistencies.
A key observation in the offset degeneracy is that for two planetary ($q\ll1$) lenses that differ only by an \textit{offset} to the projected star-planet separation ($s_A\neq s_B$) on the same lens-axis, their locus of equal magnification --- referred to as the \textit{null} --- intersects with the lens-axis at
\begin{equation}
    \nult=\dfrac{s_A-1/s_A+s_B-1/s_B}{2},
    \label{eq:xnull}
\end{equation}
where the subscript ``0'' indicates to zeroth-order in $q$, which we prove to be the correct form in Section \ref{sec:null}.
The intersection between the null and the lens-axis is referred to as the \textit{lens-axis null} hereafter as a shorthand. Given that planetary anomalies primarily occur on and near the lens-axis, source trajectories crossing the lens-axis null
\begin{equation}
\label{eq:traj}
    \dfrac{u_0}{{\rm sin}(\alpha)}=\nult
\end{equation}
are then expected to result in similar light-curves under the null-forming lens configurations.
In the above equation, $u_0/{\rm sin}(\alpha)\equiv u_{\rm anom}$ is where the source crosses the lens-axis, which is usually also the source-star separation around the midpoint of the planetary anomaly, $u_0$ is the impact parameter to the coordinate origin (see Section \ref{sec:inside} for detailed considerations), and $\alpha$ is the angle between the source trajectory and the lens axis.

Crucially, the above formalism is continuous over caustic topology transitions for $q\ll1$, and thus generalizes the close-wide and inner-outer degeneracies to the resonant regime. One major implication is that the close-wide degeneracy only strictly manifests for the singular case of $u_0=0$, and elsewhere the offset degeneracy predicts a deviation from $s\leftrightarrow 1/s$. We thus refer to the close-wide degeneracy as the central caustic degeneracy, in line with \cite{an_condition_2021}. While \cite{zhang_ubiquitous_2022} verified that the above formalism accurately describes the degenerate solutions in 23 observed events in the referred literature, it was found numerically and no theoretical justification was given. Subsequently, an alternative formalism for the unification of degeneracies was proposed in \citealt{gould_systematic_2022}, 
whose the relationship to the offset degeneracy will be discussed in Section \ref{sec:conclusion}.

In this work, we provide a mathematical treatment of the offset degeneracy. In Section \ref{sec:null}, the location of the lens-axis null is derived from the lens equation, which proves the formalism proposed in \cite{zhang_ubiquitous_2022}. In Section \ref{sec:alpha}, conditions on the source trajectory orientation is discussed. Finally, a generalized $N$-body offset degeneracy based on the superposition principle is discussed in Section \ref{sec:nbody}, whereas Section \ref{sec:conclusion} concludes our work.

\section{Derivations}
\label{sec:null}
The goal of this section is to answer the question: given two planetary lenses with the same mass-ratio ($q_{\rm A}=q_{\rm B}\ll1$) but different projected star-planet separations ($s_{\rm A}\neq s_{\rm B}$), where on the lens axis does their magnifications equal?

Let us begin by defining the lens equation. With the primary star on the origin and the planet on the real-axis at a distance $s$ from the primary, the two-body complex lens equation \citep{witt_investigation_1990} states
\begin{equation}
    \zeta=z-\dfrac{1-m}{\bar{z}}-\dfrac{m}{\bar{z}-s},
    \label{eq:lens}
\end{equation}
where $\zeta=\xi+i\eta$ and $z=z_1+i z_2$ are the complex source and image locations, $m$ is the planetary mass normalized to the total lens mass ($M_{\rm tot}$), and $s$ is the projected star-planet separation normalized to the angular Einstein radius $\theta_{E}=\sqrt{4G M_{\rm tot}/(D_{\rm rel}c^2)}$ where $D_{\rm rel}$ is the source-lens relative distance defined as $D_{\rm rel}^{-1}=D_{\rm lens}^{-1}-D_{\rm source}^{-1}$. 

\cite{witt_minimum_1995} showed that the lens equation can be transformed into a 5th-order polynomial in $z$ by substituting the conjugate of Equation \ref{eq:lens},
\begin{equation}
    \bar{z}=\bar{\zeta}+\dfrac{1-m}{z}+\dfrac{m}{z-s},
    \label{eq:lens2}
\end{equation}
back into itself, whereby conjugates in $\bar{z}$ are cleared. The resulting polynomial is
\begin{equation}
    p_{5}(z;\zeta,m,s)=\sum_{i=0}^{5} a_i(\zeta,m,s)\cdot z^i=0,
    \label{eq:lens3}
\end{equation}
where
\begin{align}
    a_0=&(1-m)^2s^2\zeta\nonumber\\
    a_1=&(1-m)s[ms-(2+s^2)\zeta+2s\zeta\bar{\zeta})]\nonumber\\
    a_2=&\zeta + 2 s^2 \zeta - ms (1+ s\zeta) \nonumber\\&- s (s -2 m s - 2 (m-2) \zeta + s^2 \zeta)\bar{\zeta} + s^2 \zeta \bar{\zeta}^2\nonumber\\
    a_3=&-s (m s + \zeta) + (-2 (m-1) s + s^3 + 2 \zeta + 2 s^2 \zeta) \bar{\zeta}\nonumber\\ &- s (s + 2 \zeta) \bar{\zeta}^2\nonumber\\
    a_4=&m s - (1 + 2 s^2 + s \zeta) \bar{\zeta} + (2 s + \zeta) \bar{\zeta}^2\nonumber\\
    a_5=&(s - \bar{\zeta}) \bar{\zeta}\nonumber.
\end{align}

The magnification of each individual image $j$ located at $z_j$ is given by the absolute value of the inverse of the Jacobian determinant of the lens equation:
\begin{align}
    \mu_j&=\dfrac{p_j}{{\rm det} J|_{z=z_j}}\\
    &=p_j\left.\left(1-\dfrac{\partial \zeta}{\partial \bar{z}}\dfrac{\overline{\partial \zeta}}{\partial \bar{z}}\right)^{-1}\right|_{z=z_j},
\end{align}
where $p_j=\pm1$ denotes the parity of the image.

\cite{witt_minimum_1995} further demonstrated how one may acquire the individual image magnifications $\mu_j$ without solving for the image locations $z_j$. Evaluating $\partial \zeta/\partial \bar{z}$ with Equation \ref{eq:lens}, clearing conjugates in $z$ with Equation \ref{eq:lens2}, and clearing fractions, one obtains a 8th-order polynomial in $z$ whose coefficients are parameterized by $\mu_j$. From here on, let us restrict our discussion to the lens-axis, i.e., the real-axis ($\zeta=\xi$). The common variable $z$ in this 8th-order polynomial and 5th order polynomial associated with the lens equation (Equation \ref{eq:lens3}) can be eliminated by calculating their resultant, which results in a lengthy 5th-order polynomial in $\mu$:
\begin{equation}
    p_5(\mu;\xi,m,s)=\sum_{i=0}^{5} b_i(\xi,m,s)\cdot \mu^i=0.
    \label{eq:mu}
\end{equation}
whose coefficients are parametrized by $\xi$, $m$, and $s$. The above polynomial can be further factored into linear and cubic polynomials:
\begin{equation}
    p_5(\mu;\xi,m,s)=\left(\sum_{i=0}^{1} c_i\cdot \mu^i\right)^2\cdot\left(\sum_{i=0}^{3} d_i\cdot \mu^i\right)=0.
    \label{eq:mu1}
\end{equation}

Of the five solutions $\mu_j$, the equal-magnification solutions ($\mu_1=\mu_2=-c_0/c_1$) for the linear equation correspond to the two off-axis images that only exist when the source is inside of a caustic and are positive in parity. The cubic polynomial has three real roots which correspond to three negative parity images ($\mu_{3,4,5}<0$) when the source is inside of caustics, but one positive and two negative parity images when the source is outside of caustics \citep{witt_minimum_1995}. Let us now consider these two cases separately.

\subsection{Inside Caustics}
\label{sec:inside}
\begin{figure}
 \centering
 \includegraphics[width=\columnwidth]{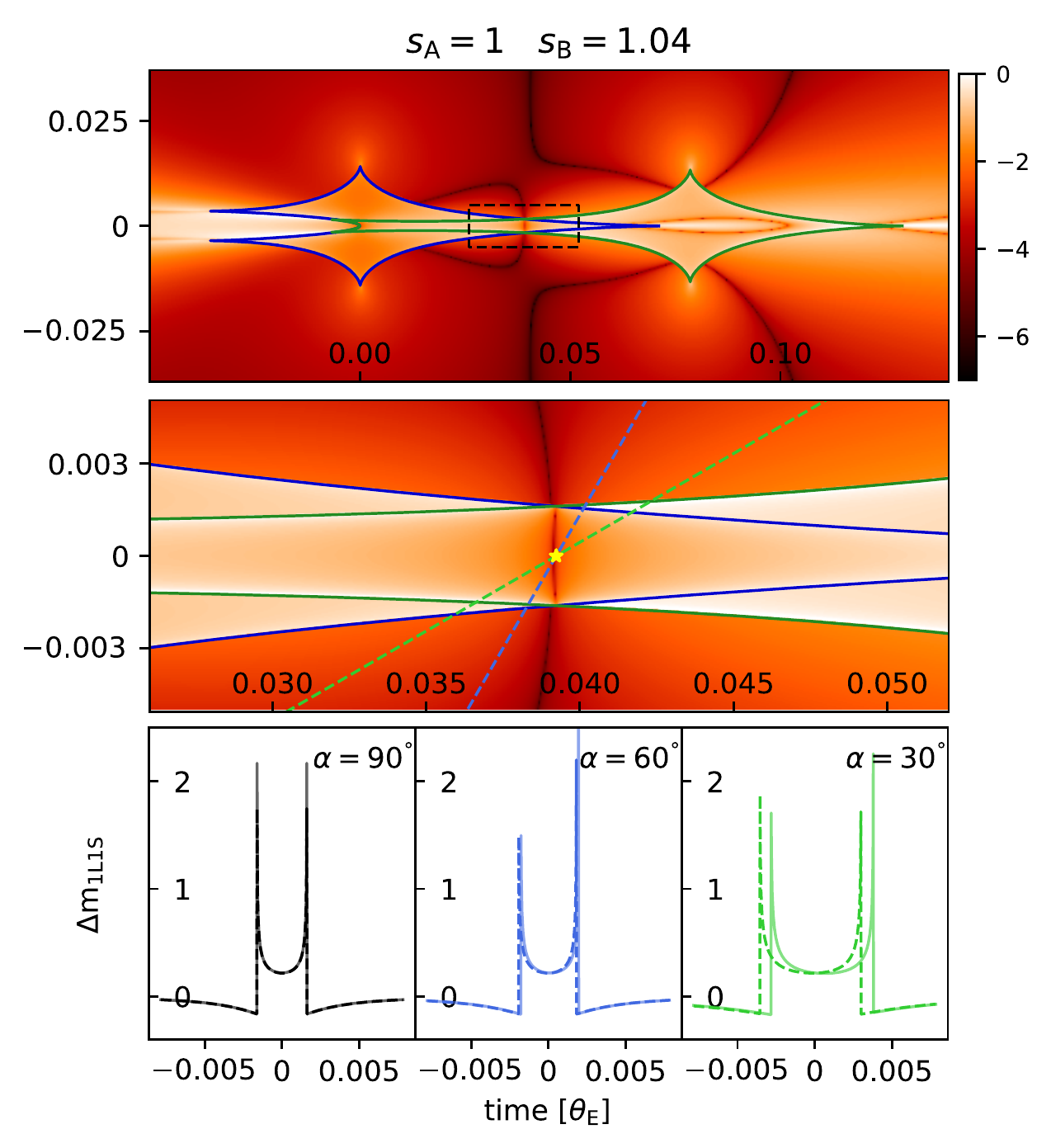}
 \caption{Top: fractional magnification difference between ($s_{\rm A}=1$, $q=10^{-4}$) and ($s_{\rm B}=1.04$, $q=10^{-4}$), with color-scale shown to the right in $\log_{10}$. Black contours illustrate the locus of equal magnification. The x and y axes are in units of $\theta_{\rm E}$. Middle: a zoom-in of the dashed-line boxed region in the top panel. The location of the lens-axis null expected from $\nult$ is marked with the gold star in the center. Source trajectories with $\alpha=30^\circ, 60^\circ$ are shown in green and blue dashed lines. Bottom: differences to single-lens light-curves for null-crossing trajectories. Dashed lines corresponds to $s_A=1$ whereas solid lines are for $s_B=1.04$. Trajectory orientation is marked in the subplot upper-right corners with the same color coding as the middle plot. The $\alpha=30^\circ$ case is seen to have different caustic entry-exit times but similar caustic-crossing durations.}
 \label{fig:in}
\end{figure}

When the lens-axis null --- the intercept of the locus of equal magnification on the lens axis --- is located inside of caustics (Figure \ref{fig:in}), images for each of the two polynomials in Equation \ref{eq:mu1} are respectively equal in parity and the total magnification can be derived directly from the polynomial coefficients:
\begin{align}
    \mu_{\rm tot,in}(\xi,m,s)=&(\mu_1+\mu_2)-(\mu_3+\mu_4+\mu_5)\nonumber\\
    =&-2c_0/c_1+d_2/d_3\nonumber\\
    \mu_{\rm tot,in}(\xi,m,s)=&\dfrac{3m^2 s^2 -\xi^2 A^2 + 2 m s B}{
 m^2 s^2 +\xi^2 A^2 - 
  2 m s \xi C},
    \label{eq:tot}
\end{align}
where
\begin{align}
    A=&1-s^2+s\xi\nonumber\\
    B=&-2 s + (1 + s^2) \xi - 3 s \xi^2 + 2 \xi^3\nonumber\\
    C=&1 + s^2 - 3 s \xi + 2 \xi^2.\nonumber
\end{align}

The location of the lens-axis null can be derived by solving $\mu_{\rm tot,in}(s_A)=\mu_{\rm tot,in}(s_B)$. Since for planetary microlenses $m\ll1$, the $m^2s^2$ term can be dropped in both the numerator and the denominator, and we can substitute the planet-to-star mass ratio $q=m/(1-m)$ for $m$. Clearing fractions in $\mu_{\rm tot,in}(s_A)-\mu_{\rm tot,in}(s_B)=0$, we obtain a quadratic polynomial in $\xi$. Taking the zeroth-order Taylor expansion in $q$, one of the roots simplifies to
\begin{equation}
    \xi_{\rm null,in}=\dfrac{s_A-1/s_A+s_B-1/s_B}{2}+\bigo{q},
    \label{eq:nullin}
\end{equation}
where the other root is reduced to 0. We have thus shown that the empirically derived $\nult$ (Equation \ref{eq:xnull}) is exact for null-in-caustic to zeroth-order in $q$.

To see how $\xi_{\rm null,in}$ may deviate from the zeroth-order term ($\nult$) for finite value of $q$, let us now consider the first-order term in $q$ and its dependence on $s_{\rm A,B}$. In particular, for $s_{\rm A}=1/s_{\rm B}$, we should expect the first-order term to not diverge to infinity in the $s_{\rm A,B}\rightarrow\{0,\infty\}$ limit, in order to be consistent with the central caustic degeneracy.
Here, it is important to adapt a coordinate origin that is consistent with caustic degeneracies. \cite{an_condition_2021} noted that while the central caustic degeneracy breaks down near the resonant regime, a pair of resonant caustics with $s_{\rm A}=1/s_{\rm B}$ still resembles each other locally towards the back end of the caustic (near the primary star). This suggests that one should choose a coordinate origin that consistently aligns the back-end of the central/resonant caustic for a pair of lenses with an arbitrary difference in separation ($s_{\rm A,B}$). 

\begin{figure}
 \centering
 \includegraphics[width=\columnwidth]{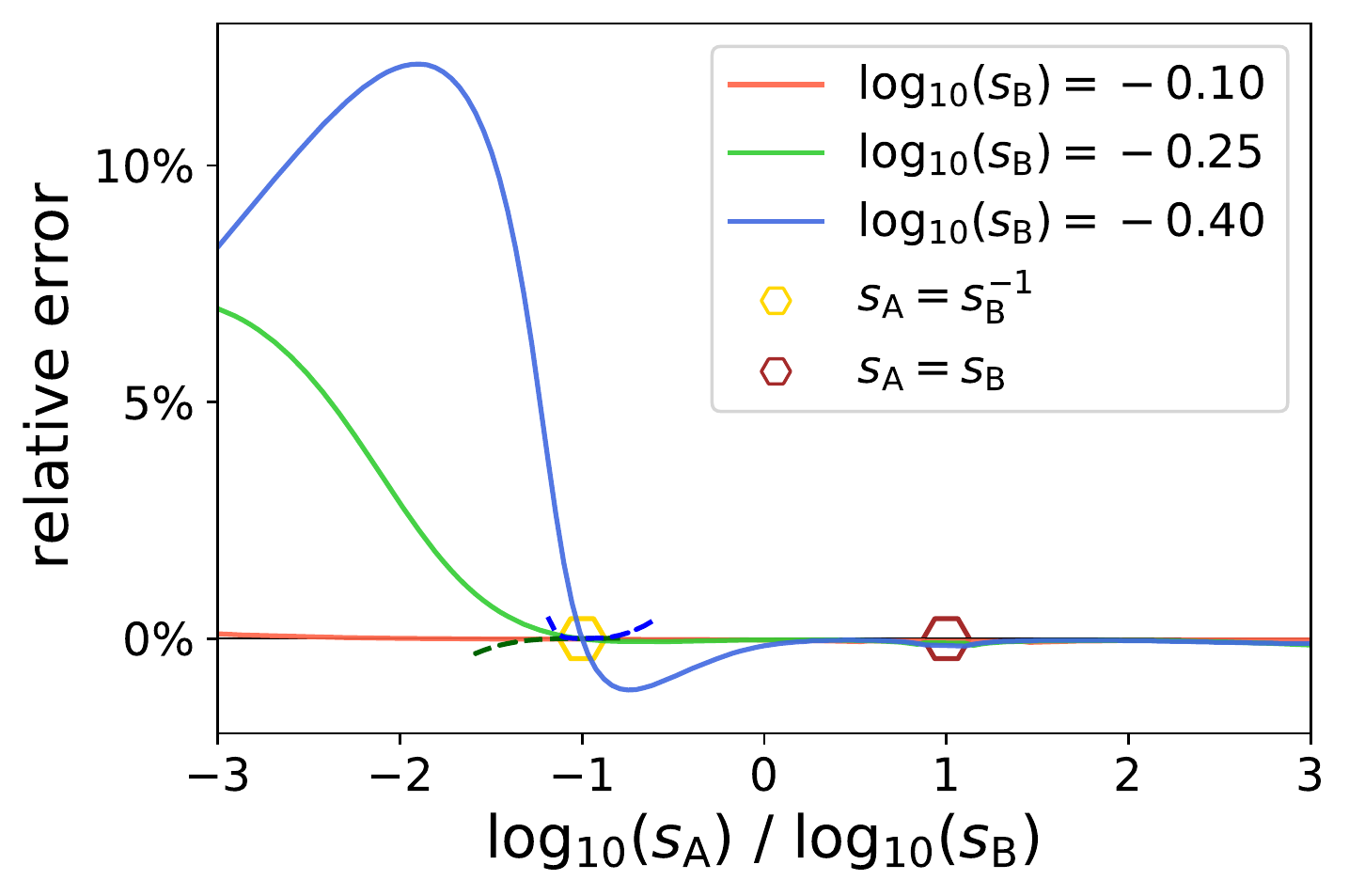}
 \caption{Deviation of $\nult$ from the exact null location, normalized to $|(s_A-1/s_A)-(s_B-1/s_B)|$, where the exact null location is derived numerically with $q=10^{-4}$. Three solid curves show this relative error for changing $s_A$ against three values of fixed $s_B\simeq(1/1.3, 1/1.8, 1/2.5)$. The two dashed lines with darker colors show the alternative expression $\xi_{\rm null,hm}$ which is exact for $\nul\ll1$ (see Section \ref{sec:out}), or equivalently $s_{\rm A}\sim1/s_{\rm B}$, shown only for $|\nul|<0.5$ and $|s_{\rm A}-s_{\rm B}|>1$.}
 \label{fig:0}
\end{figure}

We therefore opt to use the effective primary star location \citep{di_stefano_microlensing_1996,An_2002,chung_properties_2005} as the coordinate origin, which is given by
\begin{equation}
    \xi\rightarrow\xi+\dfrac{q}{(1+q)\cdot(s+s^{-1})},
    \label{eq:eff}
\end{equation}
and indeed achieves the aforementioned alignment.
Note that the effective primary location reduces to
\[
\xi\rightarrow
\begin{cases}
\xi+sq/(1+q) & s\ll1\\
\xi+s^{-1}q/(1+q) & s\gg1,\\
\end{cases}
\]
which are the central caustic locations \citep{han_distinguishing_2008} that were used in \cite{zhang_ubiquitous_2022} as the coordinate origin for their numerical calculations.
We point out that the $\sim2\%$ error at $s_{\rm A}=1$ and $s_{\rm B}=0.4$ in Figure 2 of \cite{zhang_ubiquitous_2022} is a direct result of their coordinate choice, which is inaccurate in describing resonant caustic locations and causes a misalignment between the resonant and central caustics. Figure \ref{fig:0} reproduces that same figure, but with the effective primary (Equation \ref{eq:eff}) as the origin, and shows that the error of $\nult$ at $s_{\rm A}=1$ and $s_{\rm B}=0.4$ is reduced to $0.1\%$ and remains $<0.1\%$ for $|\log(s_{\rm A,B})|<0.25$, or $1/1.8<s_{\rm A,B}<1.8$.

Applying the above coordinate transformation to the previous derivation, we find that while the zeroth-order term remains $\nult$ as expected, the first-order term ($f\cdot q$) is rather involved. There are only two special cases that are relevant here.

If the null is located within the central caustic, we should expect $s_A\sim1/s_B$, which simplifies the first order term $f\cdot q$ to
\begin{equation}
    f\sim-\dfrac{s (3 + 2 s^2 + 3 s^4)}{(1 + s^2)^3}.
\end{equation}
Note that the above expression is symmetrical under $s\leftrightarrow s^{-1}$. Since $f\rightarrow 0$ for $s\rightarrow\{0,\infty\}$, $f$ does not diverge and is typically of order unity. However, if we had defined the lens-equation (Equation \ref{eq:lens}) in units of the Einstein radius of the primary mass, then $f$ diverges to infinity for both $s\rightarrow\{0,\infty\}$, justifying our choice of parameterization with the Einstein radius of the total mass.

On the other hand, if the null is within the resonant or the wide-planetary caustic, we should expect $s_A\simeq s_B\gtrsim1$, which results in
\begin{equation}
    f\sim -\dfrac{2}{s+s^3},
\end{equation}
and is also order unity.
One may thus expect $\xi_{\rm null,in}\simeq\nult-q$, that is, a deviation of order $q$, which is in agreement with the slight deviation seen in the middle panel of Figure \ref{fig:in}.

\subsection{Outside Caustics}
\label{sec:out}
\begin{figure*}
 \centering
 \includegraphics[width=\textwidth]{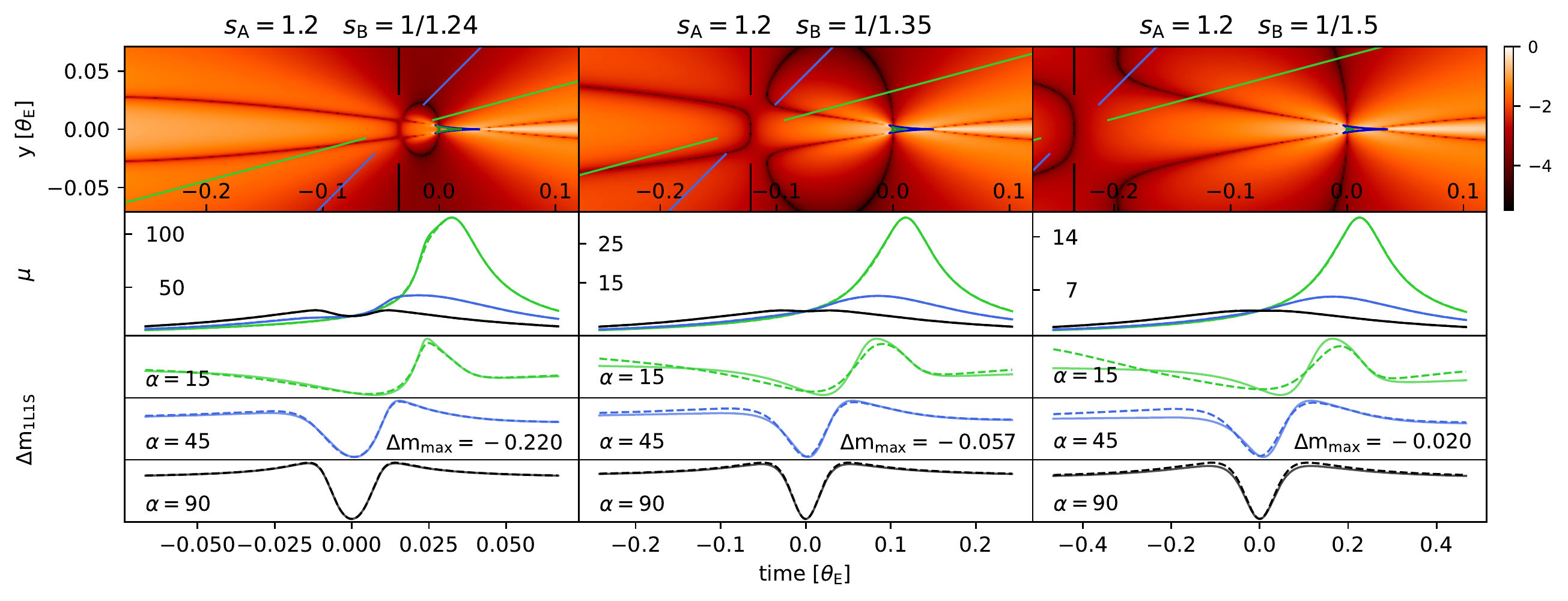}
 \caption{Top row: magnification difference in log-scale for three pairs of lens configurations indicated in the subplot titles. $q=10^{-3}$ for all cases. Color-bar to the right shows the difference scale in $\log_{10}$. The oval-shaped contours are the loci of equal magnification (null). Three null-crossing source trajectories with $\alpha=15^\circ,45^\circ,90^\circ$ are shown with the two-segment solid lines, with direction going from upper-right to lower-left. The green central caustics are for the changing $s_B$. Second row: magnifications ($\mu$) for null-crossing trajectories in the same color coding as the top row. Solid lines are for $s_A$ and dashed lines for $s_B$. The x-axis (time) is centered on the lens-axis null and scaled to $|\nul|$. Bottom three rows: planetary perturbation shown as the difference to a single lens model in unit of magnitudes. The maximum deviation is indicated in the second-to-last row.}
 \label{fig:1}
\end{figure*}

\begin{figure*}
 \centering
 \includegraphics[width=\textwidth]{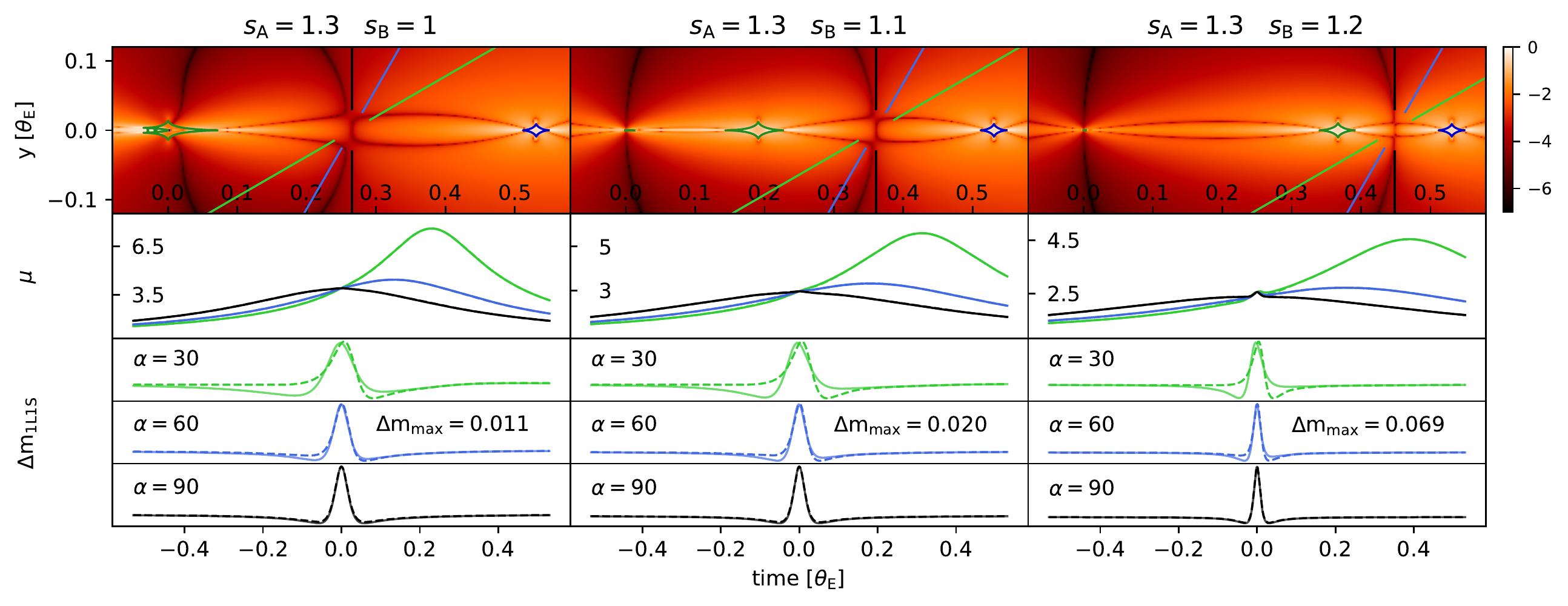}
 \caption{Same as Figure \ref{fig:1} but for three different configurations.}
 \label{fig:2}
\end{figure*}

For sources outside caustics (Figure \ref{fig:1} \& \ref{fig:2}), there are three images which are different in parity, and we can no longer obtain the total magnification directly from the polynomial coefficients. The sum of the absolute value of the cubic roots is also difficult to simplify. However, keeping coefficients up to first order in $q$, the cubic part of Equation \ref{eq:mu1} is reduced to a quadratic polynomial with two roots that are in a much simpler form compared to the cubic roots.
The total magnification is then the absolute difference between the two roots representing one positive and one negative parity image.
Indeed, when the source is away from the planetary caustic, the image closest to the planet typically has negligible magnification. As for the alternative scenario, we should already expect $\nult$ to hold in the immediate vicinity of planetary caustics, given that the location of the lens-axis null transitions continuously from inside to outside of caustics.

Equating the total magnification for $s_A$ and $s_B$, clearing fractions, further taking the first order expansion in $q$ and simplifying, we acquire a quartic polynomial
\begin{equation}
    p_{\rm null}(\xi;s_{\rm A},s_{\rm B})=\sum_{i=0}^{4} e_i(s_{\rm A},s_{\rm B})\cdot\xi^i=0,
    \label{eq:nullout}
\end{equation}
whose coefficients are provided in Appendix A. This polynomial could be solved for the lens-axis null outside of caustics for any arbitrary pair of $s_{\rm A,B}$ satisfying $q\ll1$.

To examine the conditions for $\nult$ to be the exact form to zeroth-order in $q$, let us directly plug $\nult$ into $p_{\rm null}$ as an ansatz, which reduces the polynomial to
\begin{equation}
    -\dfrac{(s_A - s_B)^6 (s_A s_B-1) (s_A s_B+1)^2}{4 s_A^2 s_B^2}=\bigo{(s_A-s_B)^6}.
\end{equation}
Given non-zero first order derivative $p'_{\rm null}$ and bounded higher order derivatives, $p_{\rm null}\rightarrow0$ implies $\xi\rightarrow\nult$, that is, the ansatz is indeed a root. Thus $\nult$ is exact for $(s_{\rm A}-s_{\rm B})^6\ll1$ to zeroth-order in $q$. Note that this condition is substantially more relaxed than the $|s_{\rm A}-s_{\rm B}|\ll1$ condition (e.g. $0.5^6\simeq0.015$). Furthermore, the condition of the lens being near the resonant regime ($|1-s|\lesssim q^{1/3}$) is a sufficient condition for $(s_{\rm A}-s_{\rm B})^6\ll1$, allowing $\nult$ to be essentially exact for semi-resonant events.

Numerical calculations (Figure \ref{fig:0}) show that the error on $\nult$ remains less than $1\%$ for $1/2.5<s_{\rm A,B}<2.5$ and should be sufficiently accurate for practical purposes. Larger deviations of a few percent are found near $s_{\rm A}\sim 1/s_{\rm B}$ where $|s_A-s_B|\gtrsim3$. As a theoretical exercise, an alternative expression for these high-magnification ($\nul\ll1$) events can be immediately acquired by linearizing $p_{\rm null}$ in $\xi_{\rm null}$, which results in:
\begin{equation}
    \xi_{\rm null, hm}=-e_0/e_1,
    \label{eq:hm}
\end{equation}
where the coefficients can be found in Appendix A.
Figure \ref{fig:0} shows $\xi_{\rm null, hm}$ for $|\nul|<0.5$ (dashed lines), which verifies that $\xi_{\rm null, hm}$ indeed describes the local behavior at $s_A\sim1/s_B$.

\section{Source trajectory orientation}
\label{sec:alpha}
Technically, the above derivation only guarantees exact magnification matching on the lens-axis. It was shown in \cite{zhang_ubiquitous_2022} that vertical null-crossing trajectories result in nearly identical light-curves, which was also noted in \cite{gaudi_planet_1997} for the inner-outer degeneracy. Indeed, Figures \ref{fig:in}, \ref{fig:1}, \ref{fig:2} all demonstrate that the locus of equal magnification is vertically extended near the lens-axis. Here, we consider the extend to which oblique trajectories could remain degenerate.

Let us first consider the case where the lens-axis null is located outside of caustics. Figure \ref{fig:1} shows three examples where the null gradually moves away from the central caustic. Figure \ref{fig:2} shows three additional cases where $s_B$ approaches $s_A$ from $s_B=1$. Note how in Figure \ref{fig:2} $|\nul|$ is greater than the examples in Figure \ref{fig:1}.
In both cases, vertical trajectories essentially give rise to identical light-curves. As the trajectory becomes more oblique, the magnifications under the two degenerate lenses begin to differ in the ``wings'' of the planetary perturbation, and thus sufficiently precise photometry can break the degeneracy. By comparing Figure \ref{fig:1} and \ref{fig:2}, one may see that the trajectory angle can be as oblique as $\alpha=15^\circ$ while the light-curves remain largely the same when the null is close to the central caustic ($|\nul|\ll1$). Elsewhere, the differences on the perturbation ``wings'' become a significant fraction of the peak planetary perturbation for $\alpha\lesssim45^\circ$. While not shown, close approaches to the off-axis cusps of the planetary caustic with oblique trajectories will decisively break the degeneracy, as the time-of-approach will be either before or after crossing the lens-axis.

For the lens-axis null inside of caustics, there is notably an additional constraint on the caustic entry-exit times and duration. Figure \ref{fig:in} illustrates how the vertical null directionality implies that the caustic height is automatically matched at the lens-axis null, allowing the caustic entry-exit times and duration to be the same for vertical null-crossing trajectories. Essentially, intersections of caustics are the set of points in the source plane where magnifications for the two lenses diverge simultaneously, and by definition, must occur on the locus of equal magnification.

For oblique trajectories, note how the two resonant caustics are approximately the reflection of one another along the vertical null (black broken line in Figure \ref{fig:in}) and appears like large planetary caustics. Because of this symmetry, the caustic-crossing duration remains approximately the same, but the caustic entry-exit times begin to differ, the extent of which depends on how quickly the caustic height changes ($\left.{\rm d} \eta_{\rm caus}/{\rm d} \xi_{\rm caus}\right|_{\xi=\xi_{\rm null,0}}$) near the lens-axis null. Fine tuning of the lensing parameters (e.g. the event timescale) may reduce the difference in the caustic entry-exit times. Additionally and similarly to null-outside-caustic, close approaches to the off-axis cusps (not shown in Figure \ref{fig:0}) will be asymmetrical for oblique trajectories would categorically break the degeneracy. Finally, for the lens-axis null inside of central caustics ($|1-s|\ll q^{1/3}$), the central caustics are close to identical due to the central caustic degeneracy and thus the aforementioned constraints on the caustic entry-exit times are less relevant.

Recent examples in the literature of null-in-caustic include, among others, KMT-2019-BLG-0371 \citep{2021AJ....162...17K}, KMT-2019-BLG-1042 \citep{zang_systematic_2022}, and OGLE-2019-BLG-0960 \citep{yee_ogle-2019-blg-0960_2021}. In the case of OGLE-2019-BLG-0960, the trajectory was quite oblique ($\alpha\simeq15$), yet still resulted in very degenerate solutions because the caustic height in this particular case changes slowly near the null ($\left|{\rm d} \eta_{\rm caus}/{\rm d} \xi_{\rm caus}\right|_{\xi=\xi_{\rm null,0}} \ll 1$), allowing the caustic entry-exit times to remain approximately the same even for very oblique trajectories.

\section{Generalization to $N$-body lens}
\label{sec:nbody}

\begin{figure}
 \centering
 \includegraphics[width=0.95\columnwidth]{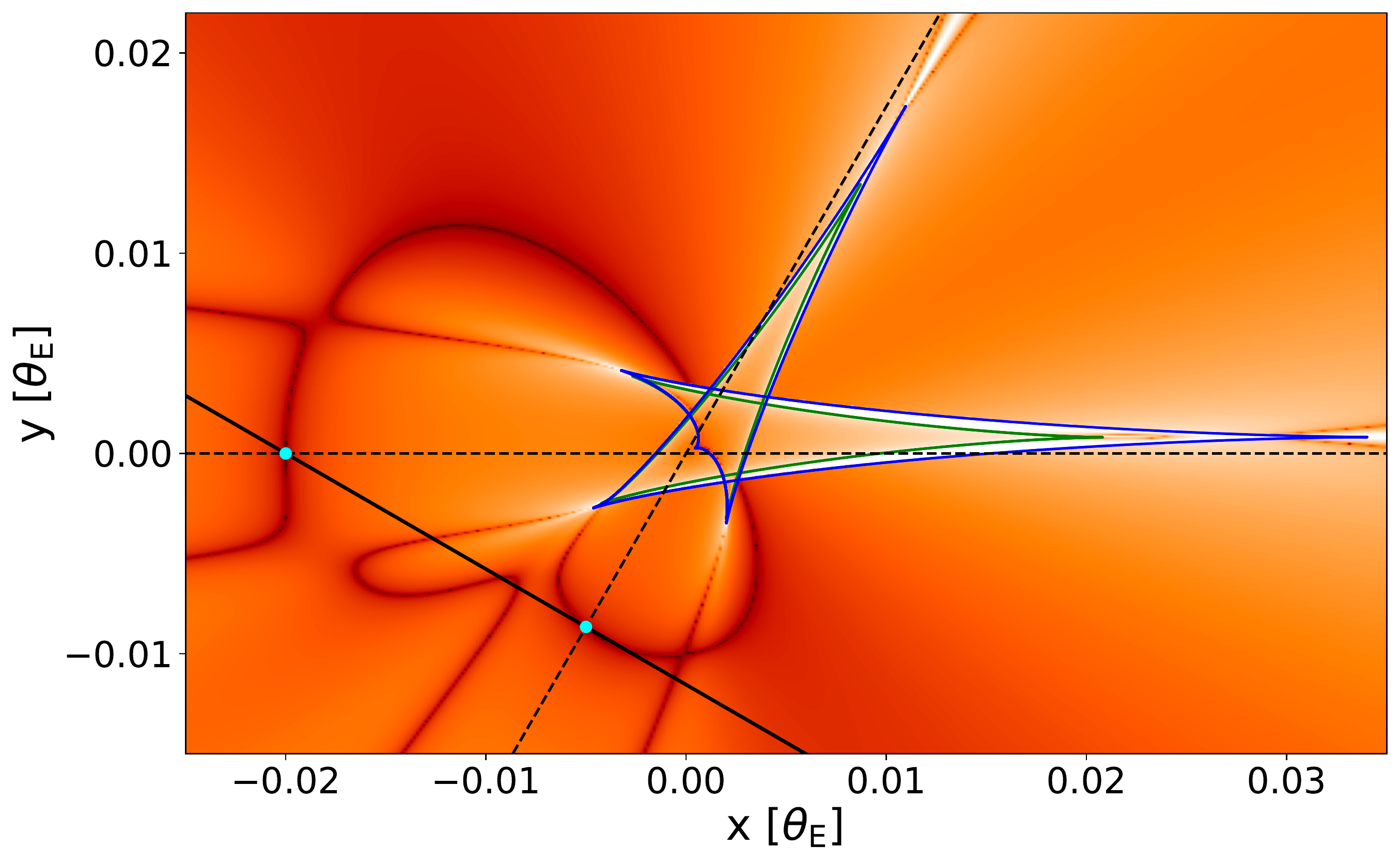}
 \includegraphics[width=\columnwidth]{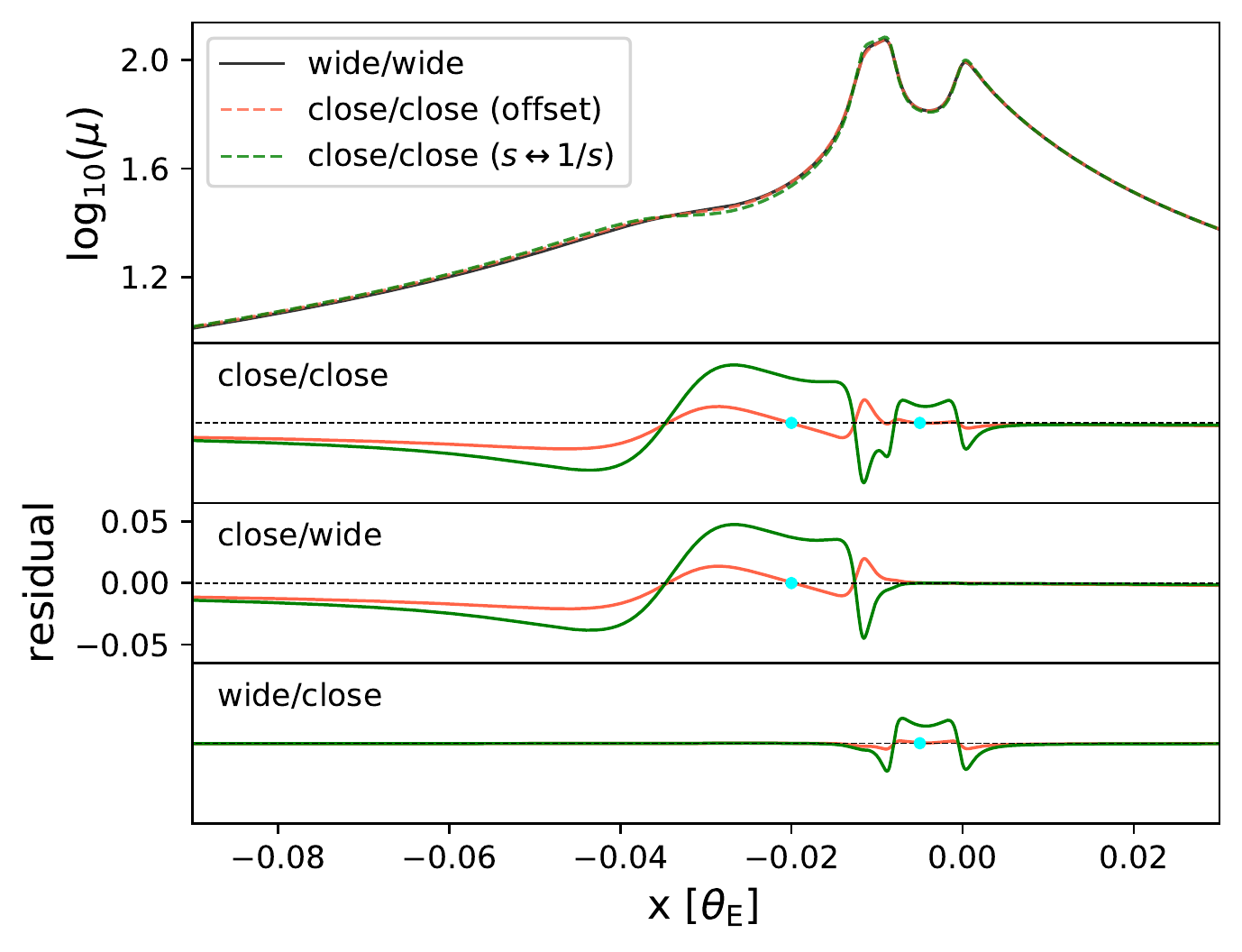}
 \caption{Example of the offset degeneracy generalized to triple lens systems. Top: magnification difference between triple lens configurations of ($s_1,s_2,\phi$)=(1.2, 1.25, 60$\degree$), referred to as the wide/wide configuration whose central caustic is shown in blue, and the close/close configuration of (0.8189, 0.7938, 60$\degree$) whose central caustic is shown in green. $\phi$ is the angle between the two lens-axes (dashed lines), with the horizontal one corresponding to $s_1$. The two resulting lens-axis nulls are marked with cyan dots, which coincide with the source trajectory (solid line). Bottom: light-curves for the null-crossing trajectory. In the legend, $s\leftrightarrow 1/s$ refers to the (1/1.2, 1/1.25, 60$\degree$) configuration expected from the central caustic degeneracy. The designations ``close'' and ``wide'' refer to the caustic topology rather than the close-wide degeneracy. The bottom panels show light-curve residuals of the degenerate configurations to the wide/wide configuration in units of magnitudes. Light-curves resulting from the central caustic degeneracy (green curves) are shown to have greater residual than that from the offset degeneracy (red curves). The horizontal axis is the source location projected to the x-axis and the cyan dots indicate the nulls allowing for a straightforward comparison to the top figure.}
 \label{fig:triple}
\end{figure}

The superposition principle \citep{bozza_perturbative_1999,han_properties_2001} states that planetary perturbations from an $N$-body lens satisfying $q_i\ll1$ is well approximated by the superposition of perturbations from each individual planet. This allows a straightforward generalization of the offset degeneracy to $N$-body lenses, which has $N-1$ number of lens-axes, and thus the number of null to match, resulting in a $2^{N-1}$ number of degenerate configurations.

Figure \ref{fig:triple} shows an example of the offset degeneracy generalized to triple lens systems, where the source passes close to the back end of the self-intersecting central caustics. We have adapted the same configuration in Figure 2 of \cite{song_degeneracies_2014} to facilitate comparison to the extension of the central caustic degeneracy to triple-lens discussed therein. The magnification difference between the wide/wide and close-close configurations is shown to be the sum of the residuals from the two singly-offset (close/wide and wide/close) configurations, which confirms the superposition picture. Additionally, as expected the 3-body offset degeneracy also serves as a correction to the 3-body central caustic degeneracy. The light-curve difference between the close/close and wide/wide configurations is greater near the null on the horizontal lens-axis ($s_1$) than the other because the source crosses the horizontal axis at $\alpha=30$ but $\alpha=90$ for the $s_2$ axis. This is in agreement with discussions in Section \ref{sec:alpha}.

Interestingly, a detailed inspection of Figure \ref{fig:triple} reveals that the central caustic cusps at the `tips' of the central caustics are actual slightly off the two lens-axes, which can be attributed to the influence of one planet on the other's caustic. This indicates that technically one may have to apply the source-null matching principle to an ``effective lens axis.'' Moreover, the superposition principle is expected to break down when the planets are close to being aligned on the same axis. Indeed, for a triple lens for which the two planets are aligned on the same axis, there is only one null that depends on the offset of both planets. We suggest that the simplest case of the axis-aligned triple planetary lens with equal mass-ratios may be analytically tractable by studying the following lens equation:
\begin{equation}
    \zeta=z-\dfrac{1-2m}{\bar{z}}-\dfrac{m}{\bar{z}-s_1}-\dfrac{m}{\bar{z}-s_2}.
    \label{eq:3lens}
\end{equation}
Details of the generalized $N$-body offset degeneracy should be explored in future work.

\section{Discussion}
\label{sec:conclusion}

In this work, we have provided a mathematical treatment of the offset degeneracy by deriving the intercept of the equal-magnification locus on the lens-axis --- the lens-axis null --- directly from the lens-equation in the limit of $q\ll1$. The numerically found $\nult$ expression \citep{zhang_ubiquitous_2022} is shown to be the exact form of the lens-axis null location inside of caustics, and outside of caustics subject to $(s_{\rm A}-s_{\rm B})^6\ll1$, to zeroth-order in $q$.
The derivations in this work demonstrate the nature of the offset degeneracy as a mathematical degeneracy deeply rooted in the lens equation itself.

The relationship between the offset degeneracy and the central caustic (close-wide) and inner-outer degeneracies has been discussed in \cite{zhang_ubiquitous_2022}. To summarize, the offset degeneracy relaxes the non-resonant ($|1-s|\gg q^{1/3}$) condition required by the two caustic degeneracies and generalizes them to a unified regime of magnification degeneracy. For sources passing close to central caustics, the offset degeneracy serves as a correction to the $s\leftrightarrow 1/s$ relationship of the central caustic degeneracy, which only strictly manifests when $u_0=0$. For this reason, we advocate that the close-wide degeneracy should be more appropriately referred to as the central caustic degeneracy (e.g.\ \citealt{an_condition_2021}), which also serves to discourage its misuse as a magnification degeneracy.

On the other hand, the inner-outer degeneracy expects the source star to pass equidistant to the planetary caustics located at $\xi_{p}=s_{\rm A,B}-1/s_{\rm A,B}$, and thus results in the same mathematical expression as the offset degeneracy. However, the Chang-Refsdal approximation to planetary caustics fails near the resonant regime \citep{dominik_binary_1999}, and thus the offset degeneracy provides a more accurate conceptual explanation. In a subsequent paper, \cite{zhang_analytic_2022} offered an alternative interpretation by showing how planetary lenses can be decomposed into Chang-Refsdal lenses with variable shear, which results in the offset degeneracy as a direct consequence. While the terms inner and outer were originally coined to refer to ``the inner[/outer] region of the planetary caustic with respect to the planet host'' \citep{han_moa-2016-blg-319lb_2018}, the idea of a generalized perturbative picture \citep{zhang_analytic_2022} suggests that they remain meaningful labels for the offset degeneracy if they refer to the lens-plane instead ---  the location of the planet being inside or outside of the image being perturbed, with respect to the primary star.

The applicability of the central caustic degeneracy to the resonant regime was previously studied in \cite{an_condition_2021}, which found that the back-end of the central/resonant caustic remains locally degenerate into the resonant regime ($|1-s|\lesssim q^{1/3}$) but the front end becomes different. They further suggested that in this case, slight adjustments to the $q_A=q_B$ and $s_A=1/s_B$ pair of solutions may result in a locally degenerate model. This work directly responds to their suggestion: $q_A=q_B$ should remain the same whilst $s_{A,B}$ should be adjusted such that the location of the lens-axis null coincides with the source trajectory. Strictly speaking, the $q_A=q_B$ condition is an assumption made in this work which is known to be true for the caustic degeneracies. The fact that vertical trajectories give rise to identical light-curves (Figures \ref{fig:in}, \ref{fig:1}, \ref{fig:2}) validates the $q_A=q_B$ assumption, but a formal proof would require examining the magnification off the lens-axis.

While examining the magnification-matching behavior on the lens-axis is a direct way of deriving the offset degeneracy formalism, there is a potential pathway to derive the $\nult$ formalism for the null-in-caustic case by studying caustic resemblances, which was proposed by \cite{an_condition_2021}. In Section \ref{sec:alpha}, we found that the caustic height for the offset-degenerate pair of lenses matches exactly at the lens-axis null, but such a claim is based on the observation that the null is vertically-directed near the lens-axis. Therefore, studying the intersection between caustics of lenses with equal mass-ratios may be not only be an independent pathway to deriving the offset degeneracy formalism, but also a verification of the equal mass-ratio condition.

\begin{figure}
 \centering
 \includegraphics[width=\columnwidth]{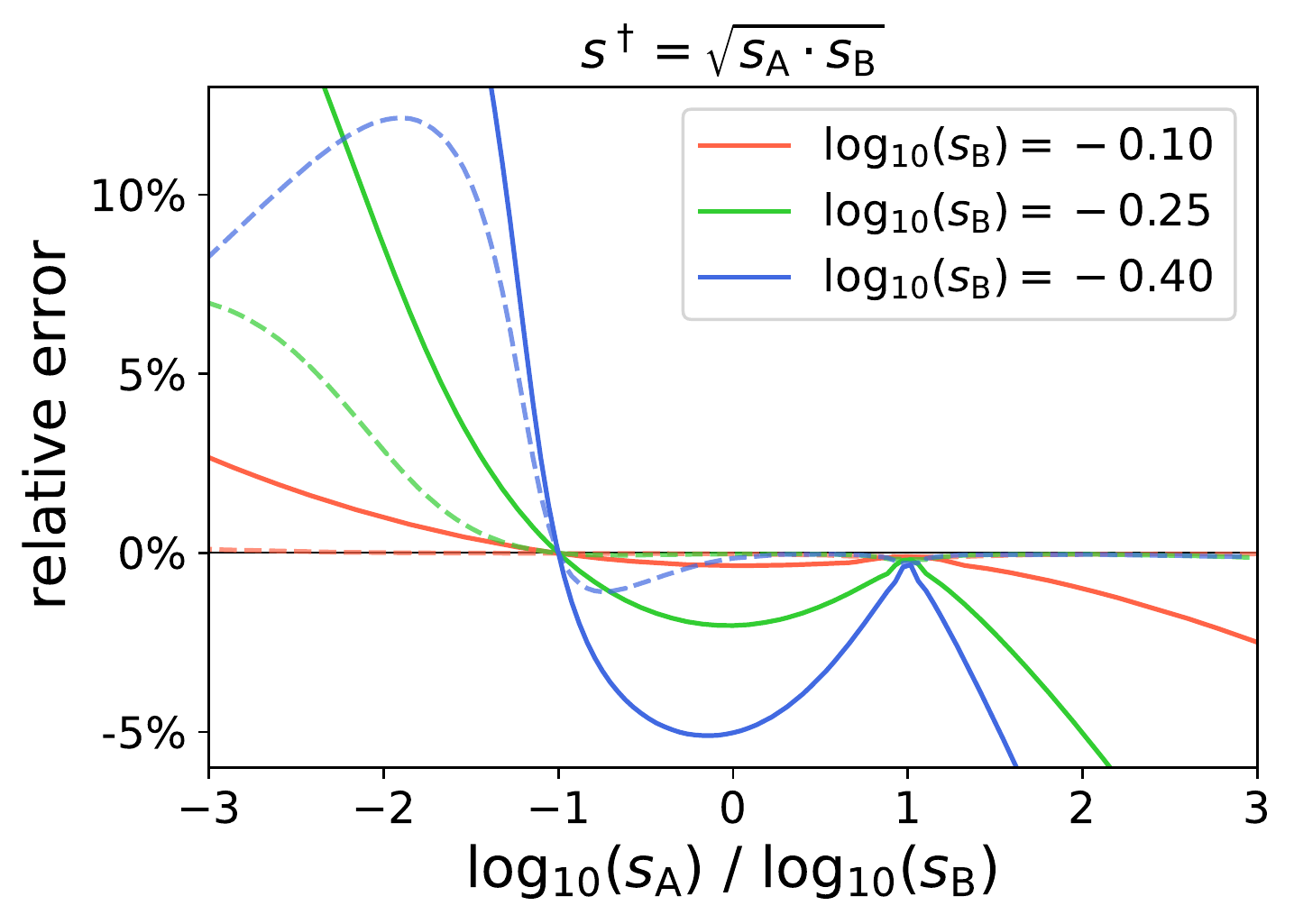}
 \caption{Error on the $s^\dagger=\sqrt{s_{\rm A}\cdot s_{\rm B}}$ heuristic, defined as the difference between the predicted value of $u_{\rm anom}= s^\dagger-1/s^\dagger$ from $s_{\rm A,B}$, and the exact location of equal magnification on the lens-axis. Solid curves are for the $s^\dagger$ heuristic and dashed curves are for the offset degeneracy ($u_{\rm anom}=\nult$) for comparison. Quantities are defined similarly to Figure \ref{fig:0}.}
 \label{fig:sdagger}
\end{figure}

Subsequent to the proposal of the offset degeneracy, \cite{ryu_mass_2022} and \cite{gould_systematic_2022} proposed an alternative formalism for unifying the close-wide and inner-outer degeneracies, referred to as the ``$s^\dagger$ heuristic''. The quantity $s^\dagger$ is defined by 
\begin{equation}
\label{eq:sdagger}
    s^\dagger=(\sqrt{u_{\rm anom}^2+4} + u_{\rm anom})/2,
\end{equation}
which is a solution to $u_{\rm anom}=s^\dagger-1/s^\dagger$, and thus the solution for planetary-caustic-crossing events. Here, we have defined $u_{\rm anom}$ as the \textit{signed} location of where the source crosses the binary axis to avoid a sign ambiguity in the original expression. This quantity was initially used in \cite{hwang_systematic_2022} for the heuristic analysis of events subject to the inner-outer degeneracy, where the solutions are approximately related by $s_{\rm A,B}=s^\dagger\pm\Delta s$. More recently, \cite{gould_systematic_2022} proposed that an alternative expression, $s^\dagger=\sqrt{s_{\rm A}\cdot s_{\rm B}}$, would lead to the unification of the two degeneracies.

The derivations in this work show that the $s^\dagger=\sqrt{s_{\rm A}\cdot s_{\rm B}}$ expression does not correctly unify the close-wide and inner-outer degeneracies, but nevertheless provides approximate solutions in the $s\rightarrow1$ limit. By substituting $\nult$ for $u_{\rm anom}$ in Equation \ref{eq:sdagger}, we find that the first order Taylor expansion of $(s^\dagger)^2$ at $s_{\rm A,B}=1$ is indeed $s_{\rm A}\cdot s_{\rm B}$. Figure \ref{fig:sdagger} shows that although the $s^\dagger=\sqrt{s_{\rm A}\cdot s_{\rm B}}$ heuristic captures the boundary cases of $s_{\rm A}=1/s_{\rm B}$ with $s^\dagger=1$ (and $u_{\rm anom}=0$), and $s_{\rm A}=s_{\rm B}=s^\dagger$, it is only approximately correct in the intermediate regime. Lastly, we note that both the $s^\dagger$ heuristic and the offset degeneracy formalism require solving one quadratic equation to derive one solution from the other based on the source trajectory, which indicates that the exact form given by Equation \ref{eq:xnull} \& \ref{eq:traj} is equally convenient to use for heuristic analysis.

\bigskip
\acknowledgements{K.Z.\ is supported by a Gordon and Betty Moore Foundation Data-Driven Discovery grant. K.Z.\ thanks the LSSTC Data Science Fellowship Program, which is funded by LSSTC, NSF Cybertraining Grant \#1829740, the Brinson Foundation, and the Moore Foundation; his participation in the program has benefited this work. Work by B.S.G.\ is supported by NASA grant NNG16PJ32C and the
Thomas Jefferson Chair for Discovery and Space Exploration. We thank Joshua Bloom, Shude Mao, and Jin An for helpful discussions, and Joshua Bloom and Shude Mao for comments on a draft of this paper.}

\software{
    \texttt{MulensModel} \citep{poleski_modeling_2019},
    \texttt{triplelens} \citep{kuang_light_2021}.
}

\newpage

\appendix
\section{Polynomial Coefficients}
Equation \ref{eq:nullout}:
\begin{align}
    e_0=&-16 (s_A s_B-1) (s_A^2 + s_A s_B + s_A^3 s_B + s_B^2 + s_A^4 s_B^2 + s_A s_B^3 + s_A^3 s_B^3 + s_A^2 s_B^4)\nonumber\\
    e_1=&-2 (s_A + s_B) (3 - 4 s_A^2 + s_A^4 - 16 s_A s_B - 4 s_B^2 + 8 s_A^2 s_B^2 - 4 s_A^4 s_B^2 - 
 16 s_A^3 s_B^3 + s_B^4 - 4 s_A^2 s_B^4 + 3 s_A^4 s_B^4)\nonumber\\
    e_2=&-4 (s_A s_B-1) (s_A^4 - 3 s_A s_B + 5 s_A^3 s_B + 6 s_A^2 s_B^2 + 5 s_A s_B^3 - 3 s_A^3 s_B^3 + s_B^4)\nonumber\\
    e_3=&-(s_A + s_B) (1 + s_A^2 - 8 s_A^3 s_B + s_B^2 - 14 s_A^2 s_B^2 + s_A^4 s_B^2 - 8 s_A s_B^3 + s_A^2 s_B^4 + s_A^4 s_B^4)\nonumber\\
    e_4=&2 s_A s_B (s_A s_B-1)(1 + s_A^2 + 2 s_A s_B + s_B^2 + s_A^2 s_B^2)\nonumber.
\end{align}


\begin{thebibliography}{}
\expandafter\ifx\csname natexlab\endcsname\relax\def\natexlab#1{#1}\fi
\providecommand{\url}[1]{\href{#1}{#1}}

\bibitem[{An(2021)}]{an_condition_2021}
An, J. 2021, arXiv:2102.07950 [astro-ph], arXiv: 2102.07950.
\newblock \url{http://arxiv.org/abs/2102.07950}

\bibitem[{An(2005)}]{an_gravitational_2005}
An, J.~H. 2005, Monthly Notices of the Royal Astronomical Society, 356, 1409,
  aDS Bibcode: 2005MNRAS.356.1409A.
\newblock \url{https://ui.adsabs.harvard.edu/abs/2005MNRAS.356.1409A}

\bibitem[{An \& Han(2002)}]{An_2002}
An, J.~H., \& Han, C. 2002, The Astrophysical Journal, 573, 351.
\newblock \url{https://doi.org/10.1086/340557}

\bibitem[{Bozza(1999)}]{bozza_perturbative_1999}
Bozza, V. 1999, Astronomy and Astrophysics, 348, 311.
\newblock \url{http://adsabs.harvard.edu/abs/1999A%26A...348..311B}

\bibitem[{Chang \& Refsdal(1984)}]{chang_star_1984}
Chang, K., \& Refsdal, S. 1984, Astronomy and Astrophysics, 132, 168.
\newblock \url{https://ui.adsabs.harvard.edu/abs/1984A&A...132..168C/abstract}

\bibitem[{Chung {et~al.}(2005)Chung, Han, Park, Kim, Kang, Ryu, Kim, Jeon, Lee,
  Chang, Lee, \& Kang}]{chung_properties_2005}
Chung, S.-J., Han, C., Park, B.-G., {et~al.} 2005, The Astrophysical Journal,
  630, 535, publisher: IOP Publishing.
\newblock \url{https://iopscience.iop.org/article/10.1086/432048/meta}

\bibitem[{Di~Stefano \& Mao(1996)}]{di_stefano_microlensing_1996}
Di~Stefano, R., \& Mao, S. 1996, The Astrophysical Journal, 457, 93, aDS
  Bibcode: 1996ApJ...457...93D.
\newblock \url{https://ui.adsabs.harvard.edu/abs/1996ApJ...457...93D}

\bibitem[{Dominik(1999)}]{dominik_binary_1999}
Dominik, M. 1999, Astronomy and Astrophysics, 349, 108.
\newblock \url{http://adsabs.harvard.edu/abs/1999A%26A...349..108D}

\bibitem[{Gaudi \& Gould(1997)}]{gaudi_planet_1997}
Gaudi, B.~S., \& Gould, A. 1997, The Astrophysical Journal, 486, 85, publisher:
  IOP Publishing.
\newblock \url{https://iopscience.iop.org/article/10.1086/304491/meta}

\bibitem[{Gould {et~al.}(2022)Gould, Han, Zang, Yang, Hwang, Udalski, Bond,
  Albrow, Chung, Jung, Ryu, Shin, Shvartzvald, Yee, Cha, Kim, Kim, Kim, Lee,
  Lee, Lee, Park, Pogge, Mróz, Szymański, Skowron, Poleski, Soszyński,
  Pietrukowicz, Kozłowski, Ulaczyk, Rybicki, Iwanek, Wrona, Abe, Barry,
  Bennett, Bhattacharya, Fujii, Fukui, Hirao, Silva, Kirikawa, Kondo,
  Koshimoto, Matsubara, Matsumoto, Miyazaki, Muraki, Okamura, Olmschenk, Ranc,
  Rattenbury, Satoh, Sumi, Suzuki, Toda, Tristram, Vandorou, Yama, Beichman,
  Bryden, Novati, Gaudi, Henderson, Penny, Jacklin, \&
  Stassun}]{gould_systematic_2022}
Gould, A., Han, C., Zang, W., {et~al.} 2022, doi:10.48550/arXiv.2204.04354.
\newblock \url{https://arxiv.org/abs/2204.04354v1}

\bibitem[{Griest \& Safizadeh(1998)}]{griest_use_1998}
Griest, K., \& Safizadeh, N. 1998, The Astrophysical Journal, 500, 37,
  publisher: IOP Publishing.
\newblock \url{https://iopscience.iop.org/article/10.1086/305729/meta}

\bibitem[{Han(2008)}]{han_distinguishing_2008}
Han, C. 2008, The Astrophysical Journal, 691, L9, publisher: IOP Publishing.
\newblock \url{https://doi.org/10.1088%2F0004-637x%2F691%2F1%2Fl9}

\bibitem[{Han {et~al.}(2001)Han, Chang, An, \& Chang}]{han_properties_2001}
Han, C., Chang, H.-Y., An, J.~H., \& Chang, K. 2001, Monthly Notices of the
  Royal Astronomical Society, 328, 986.
\newblock \url{https://doi.org/10.1046/j.1365-8711.2001.04973.x}

\bibitem[{Han {et~al.}(2018)Han, Bond, Gould, Albrow, Chung, Jung, Hwang, Lee,
  Ryu, Shin, Shvartzvald, Yee, Cha, Kim, Kim, Kim, Lee, Lee, Park, Pogge, Kim,
  Abe, Barry, Bennett, Bhattacharya, Donachie, Fukui, Hirao, Itow, Kawasaki,
  Kondo, Koshimoto, Li, Matsubara, Muraki, Miyazaki, Nagakane, Ranc,
  Rattenbury, Suematsu, Sullivan, Sumi, Suzuki, Tristram, Yonehara, \&
  {and}}]{han_moa-2016-blg-319lb_2018}
Han, C., Bond, I.~A., Gould, A., {et~al.} 2018, The Astronomical Journal, 156,
  226, publisher: American Astronomical Society.
\newblock \url{https://doi.org/10.3847/1538-3881/aae38e}

\bibitem[{Hwang {et~al.}(2022)Hwang, Zang, Gould, Udalski, Bond, Yang, Mao,
  Albrow, Chung, Han, Jung, Ryu, Shin, Shvartzvald, Yee, Cha, Kim, Kim, Kim,
  Lee, Lee, Lee, Park, Pogge, Mróz, Poleski, Skowron, Szymański, Soszyński,
  Pietrukowicz, Koz{\textbackslash}lowski, Ulaczyk, Rybicki, Iwanek, Wrona,
  Gromadzki, Abe, Barry, Bennett, Bhattacharya, Fujii, Fukui, Hirao, Itow,
  Kirikawa, Kondo, Koshimoto, Munford, Matsubara, Miyazaki, Muraki, Olmschenk,
  Ranc, Rattenbury, Satoh, Shoji, Silva, Sumi, Suzuki, Tristram, Yonehara,
  Zhang, Zhu, Penny, \& Fouqué}]{hwang_systematic_2022}
Hwang, K.-H., Zang, W., Gould, A., {et~al.} 2022, The Astronomical Journal,
  163, 43, publisher: American Astronomical Society.
\newblock \url{https://doi.org/10.3847/1538-3881/ac38ad}

\bibitem[{{Kim} {et~al.}(2021){Kim}, {Chung}, {Yee}, {Udalski}, {Bond}, {Jung},
  {Gould}, {Albrow}, {Han}, {Hwang}, {Ryu}, {Shin}, {Shvartzvald}, {Zang},
  {Cha}, {Kim}, {Kim}, {Kim}, {Lee}, {Lee}, {Lee}, {Park}, {Pogge}, {KMTNet
  Collaboration}, {Poleski}, {Mr{\'o}z}, {Skowron}, {Szyma{\'n}ski},
  {Soszy{\'n}ski}, {Pietrukowicz}, {Koz{\l}owski}, {Ulaczyk}, {Rybicki},
  {Iwanek}, {Wrona}, {Gromadzki}, {OGLE Collaboration}, {Abe}, {Barry},
  {Bennett}, {Bhattacharya}, {Donachie}, {Fujii}, {Fukui}, {Itow}, {Hirao},
  {Kirikawa}, {Kondo}, {Koshimoto}, {Matsubara}, {Muraki}, {Miyazaki}, {Ranc},
  {Rattenbury}, {Satoh}, {Shoji}, {Sumi}, {Suzuki}, {Tristram}, {Tanaka},
  {Yamawaki}, {Yonehara}, \& {MOA Collaboration}}]{2021AJ....162...17K}
{Kim}, Y.~H., {Chung}, S.-J., {Yee}, J.~C., {et~al.} 2021, \aj, 162, 17

\bibitem[{Kuang {et~al.}(2021)Kuang, Mao, Wang, Zang, \&
  Long}]{kuang_light_2021}
Kuang, R., Mao, S., Wang, T., Zang, W., \& Long, R.~J. 2021, Monthly Notices of
  the Royal Astronomical Society, stab509.
\newblock
  \url{https://academic.oup.com/mnras/advance-article/doi/10.1093/mnras/stab509/6149166}

\bibitem[{Poleski \& Yee(2019)}]{poleski_modeling_2019}
Poleski, R., \& Yee, J.~C. 2019, Astronomy and Computing, 26, 35.
\newblock
  \url{http://www.sciencedirect.com/science/article/pii/S221313371830026X}

\bibitem[{Ryu {et~al.}(2022)Ryu, Jung, Yang, Gould, Albrow, Chung, Han, Hwang,
  Shin, Shvartzvald, Yee, Zang, Cha, Kim, Kim, Kim, Lee, Lee, Lee, Park, \&
  Pogge}]{ryu_mass_2022}
Ryu, Y.-H., Jung, Y.~K., Yang, H., {et~al.} 2022.
\newblock \url{https://arxiv.org/abs/2202.03022v1}

\bibitem[{Song {et~al.}(2014)Song, Mao, \& An}]{song_degeneracies_2014}
Song, Y.-Y., Mao, S., \& An, J.~H. 2014, Monthly Notices of the Royal
  Astronomical Society, 437, 4006.
\newblock \url{https://doi.org/10.1093/mnras/stt2222}

\bibitem[{Witt(1990)}]{witt_investigation_1990}
Witt, H.~J. 1990, Astronomy and Astrophysics, 236, 311.
\newblock \url{http://adsabs.harvard.edu/abs/1990A%26A...236..311W}

\bibitem[{Witt \& Mao(1995)}]{witt_minimum_1995}
Witt, H.~J., \& Mao, S. 1995, The Astrophysical Journal Letters, 447, L105.
\newblock \url{http://adsabs.harvard.edu/abs/1995ApJ...447L.105W}

\bibitem[{Yee {et~al.}(2021)Yee, Zang, Udalski, Ryu, Green, Hennerley, Marmont,
  Sumi, Mao, Gromadzki, Mróz, Skowron, Poleski, Szymański, Soszyński,
  Pietrukowicz, Koz{\textbackslash}lowski, Ulaczyk, Rybicki, Iwanek, Wrona,
  Albrow, Chung, Gould, Han, Hwang, Jung, Kim, Shin, Shvartzvald, Cha, Kim,
  Kim, Lee, Lee, Lee, Park, Pogge, Bachelet, Christie, Hundertmark, Maoz,
  McCormick, Natusch, Penny, Street, Tsapras, Beichman, Bryden, Novati, Carey,
  Gaudi, Henderson, Johnson, Zhu, Bond, Abe, Barry, Bennett, Bhattacharya,
  Donachie, Fujii, Fukui, Hirao, Silva, Itow, Kirikawa, Kondo, Koshimoto, Li,
  Matsubara, Muraki, Miyazaki, Olmschenk, Ranc, Rattenbury, Satoh, Shoji,
  Suzuki, Tanaka, Tristram, Yamawaki, \&
  Yonehara}]{yee_ogle-2019-blg-0960_2021}
Yee, J.~C., Zang, W., Udalski, A., {et~al.} 2021, The Astronomical Journal,
  162, 180, publisher: American Astronomical Society.
\newblock \url{https://doi.org/10.3847/1538-3881/ac1582}

\bibitem[{Zang {et~al.}(2022)Zang, Yang, Han, Lee, Udalski, Gould, Mao, Zhang,
  Zhu, Albrow, Chung, Hwang, Jung, Ryu, Shin, Shvartzvald, Yee, Cha, Kim, Kim,
  Kim, Lee, Lee, Park, Mróz, Skowron, Poleski, Szymański, Soszyński,
  Pietrukowicz, Kozłowski, Ulaczyk, Rybicki, Iwanek, Wrona, \&
  Gromadzki}]{zang_systematic_2022}
Zang, W., Yang, H., Han, C., {et~al.} 2022, doi:10.48550/arXiv.2204.02017.
\newblock \url{https://arxiv.org/abs/2204.02017v1}

\bibitem[{Zhang(2022)}]{zhang_analytic_2022}
Zhang, K. 2022, Analytic {Simplifications} to {Planetary} {Microlensing} under
  the {Generalized} {Perturbative} {Picture},  arXiv, arXiv:2207.12412
  [astro-ph].
\newblock \url{http://arxiv.org/abs/2207.12412}

\bibitem[{Zhang {et~al.}(2022)Zhang, Gaudi, \& Bloom}]{zhang_ubiquitous_2022}
Zhang, K., Gaudi, B.~S., \& Bloom, J.~S. 2022, Nature Astronomy, 6, 782.
\newblock \url{https://www.nature.com/articles/s41550-022-01671-6}

\end{thebibliography}

\end{document}